\documentclass[11pt,twoside]{article}
\usepackage{acta-info}
\usepackage{euler}
\usepackage{amsmath}
\usepackage{amsfonts}
\usepackage{graphicx}
\usepackage{slashbox}
\usepackage{listings}
\usepackage{epstopdf}

\newcommand{\vol}{\textbf{3,} 2 (2011) 172--191}   
\newcommand{\amss}{\amssubj{%
65F60 11Y55
}}     
\newcommand{\CRs}{\CRsubj{%
B.2.4
}}   
\newcommand{\keyw}{\keywords{%
pseudo random number generators, linear recurring sequences, uniform distribution, 
matrix exponentiation, parallel arithmetic, FPGA design, hardware acceleration of computations, hardware implementation of computations
}}

\begin{document}
\title{
Modular exponentiation of matrices on FPGA-s
}
\maketitle


\twoauthors{%
\href{http://www.inf.unideb.hu/~herendi/}{Tam\'as HERENDI}
}{%
\href{http://www.inf.unideb.hu/}{University of Debrecen}
}{%
\href{mailto:herendi.tamas@inf.unideb.hu}{herendi.tamas@inf.unideb.hu}
}{%
Roland S\'andor MAJOR
}{%
\href{http://www.inf.unideb.hu/}{University of Debrecen}
}{%
\href{mailto:mroland@digikabel.hu}{mroland@digikabel.hu}
}


\short{%
T. Herendi, R. Major
}{%
Modular exponentiation of matrices on FPGA-s
}

\begin{abstract}
We describe an efficient FPGA implementation for the exponentiation of large matrices. The research is related to an algorithm for constructing uniformly distributed linear recurring sequences. The design utilizes the special properties of both the FPGA and the used matrices to achieve a very significant speedup compared to traditional architectures.
\end{abstract}


\section{Introduction}

Field-programmable gate arrays (FPGA) offer a number of special options in computation. Utilizing the unique properties of an FPGA, some algorithms that are impractical to implement on a more traditional architecture can become both convenient to create and resource-efficient. The programmable array of look-up tables commonly found on an FPGA provide both flexibility in creating logic to suit specific needs and naturally lend themselves to great parallelism in computations.

Fast operations on matrices are of great practical interest. Ways to speed up certain matrix calculations still find their way into numerous applications.

Faster implementations of matrix algorithms can be achieved either from a ``software'' point of view, by improving upon the algorithm itself, or from a ``hardware'' point of view, by using faster or differently structured architectures.

Theoretical improvements on matrix algorithms include Strassen's algorithm \cite{c12} and the Coppersmith-Winograd algorithm \cite{c2}. The naive algorithm for matrix multiplication is a well-known $\Theta(n^3)$ algorithm. Strassen's algorithm uses an idea similar to the Karatsuba-multiplication. It has a time complexity of $O(n^{\lg  7})$ by dividing the matrices into sub-matrices. Then by multiplying them in a different arrangement, it manages an overall lower multiplication count compared to the classical algorithm. Research implementing it on the Cell Broadband Engine can be found in \cite{c5}. Strassen's algorithm and its applicability to the project is briefly discussed in Section 7. The Coppersmith-Winograd algorithm further improves the complexity to $O(n^{2.376})$ by combining the idea of Strassen with the Salem-Spencer theorem. \cite{c9} discusses and compares the performance of implementations of these algorithms.

Numerous research has been done on creating efficient realizations of different matrix operations on different architectures. \cite{c8} and \cite{c10} both use FPGAs to perform matrix inversion.

The design presented here is an implementation of matrix multiplication on an FPGA. Works of similar nature can be found in \cite{c1} and \cite{c4}, dealing with FPGA configurations used for floating point matrix multiplication. \cite{c11} uses an FPGA design for digital signal processing. \cite{c3} discusses another FPGA implementation for accelerating matrix multiplication.

The research in this paper is related to an algorithm for the construction of pseudo random number generators. It requires the exponentiation of large matrices to an extremely high power. This allows for numerous optimizations to be made on the FPGA implementation, resulting in an extremely fast design. A speedup factor of  $\sim$200 is achieved compared to a highly optimized program on a more traditional architecture.

We give the details of a design implemented on a Virtex-5 XC5VLX110T FPGA that multiplies two $896 \times 896$ sized matrices. The matrices are defined over the mod 4 residue class ring. Using this property and the fact that the hardware uses 6-LUTs (Lookup Tables), we describe first a module that computes the dot product of vectors taken from $\mathbb{Z}_4^{28}$ in a single clock cycle at 100MHz clock speed. With these modules we construct a matrix multiplier module that computes the $C \in \mathbb{Z}_4^{20 \times 20}$ product matrix of $A \in \mathbb{Z}_4^{20 \times 28d}$ and $B \in \mathbb{Z}_4^{28d \times 20}$ in $d$ clock cycles at 100MHz. The significance of the value 28 in the implementation and its experimental determination is also discussed. Finally, we describe how to use these modules for multiplying matrices taken from $\mathbb{Z}_4^{896 \times 896}$. The proposed algorithm deals with the management of stored data in such a way that it can be accomplished completely in parallel with the computations. The resulting design completes the multiplication in 64800 clock cycles at 100MHz.

Future work for increasing the size of the used matrices, and further optimizing the design's performance using Strassen's algorithm is also described.

\section{Mathematical background}

The present work is initiated by a method for the construction of uniformly distributed pseudo random number generators. (See \cite{c7}.) The generator uses recurring sequences modulo powers of 2 of the form
\begin{align*}
u_n \equiv a_{d-1}u_{n-1}+a_{d-2}u_{n-2}+\cdots+a_{0}u_{n-d} \mod{2^s}, \  a_i \in \{0,1,2,3\}, s \in \mathbb{Z}^+
\end{align*}
The theoretical background can be found in \cite{c6}.

The construction assumes that the values $a_0,a_1,\ldots,a_{d-1}$ are such that
\[
x^{d}-a_{d-1}x^{d-1}-\cdots-a_0 \equiv (x-1)^{2}P(x) \mod{2}
\]
holds for some $P(x)$ irreducible polynomial. It is practical to choose $P(x)$ to have maximal order, since the order of $P$ is closely related to the period length of the corresponding recurring sequence.
The sequence $u_n$ obtained this way does not necessarily have uniform distribution, however exactly one of the following four sequences does:
\begin{align*}
&u_n^{(0)}\equiv a_{d-1}u_{n-1}^{(0)}+a_{d-2}u_{n-2}^{(0)}+\cdots+a_{1}u_{n-d+1}^{(0)}+a_{0}u_{n-d}^{(0)} \mod{2^s}\\
&u_n^{(1)}\equiv a_{d-1}u_{n-1}^{(1)}+a_{d-2}u_{n-2}^{(1)}+\cdots+a_{1}u_{n-d+1}^{(1)}+(a_{0}+2)u_{n-d}^{(1)} \mod{2^s}\\
&u_n^{(2)}\equiv a_{d-1}u_{n-1}^{(2)}+a_{d-2}u_{n-2}^{(2)}+\cdots+(a_{1}+2)u_{n-d+1}^{(2)}+a_{0}u_{n-d}^{(2)} \mod{2^s}\\
&u_n^{(3)}\equiv a_{d-1}u_{n-1}^{(3)}+a_{d-2}u_{n-2}^{(3)}+\cdots+(a_{1}+2)u_{n-d+1}^{(3)}+(a_{0}+2)u_{n-d}^{(3)} \mod{2^s} \ .\\
\end{align*}
For the details see \cite{c7}. Finding the sequence with uniform distribution is of interest. 
Let

\begin{align*}
M(u)=
\left(
\begin{matrix}
0 & 1 & \hdots & 0 & 0 \\
\vdots & \vdots & \ddots & \vdots & \vdots \\
0 & 0 & \hdots & 1 & 0 \\
0 & 0 & \hdots & 0 & 1 \\
a_0 & a_1 & \hdots & a_{d-2} & a_{d-1}
\end{matrix}
\right)
\end{align*}
be the companion matrix of sequence $u$. To find which of the above sequences has a uniform distribution, we have to compute $M(u)^{2^{d+1}-2}\mod 4$. If $M(u)^{2^{d+1}-2}\mod 4$ equals the identity matrix, then the period length of $u_n$ is $2^{d+1}-2$, which means it is not the sequence we are searching for.

The exponentiation of matrices to high powers can quickly become time consuming on traditional computers. The aim of the project was to utilize the special properties of an FPGA to achieve a significant upgrade in speed compared to implementations on more traditional architectures.

\section{Hardware used in the implementation}

The project was implemented on a Xilinx XUPV505-LX110T development platform. The board features a variety of ports for communication with the device. As a first approach the RS-232 serial port was used to send data between the board and a PC. A high-speed PCI Express connection is also available if the amount of data transferred would necessitate its use.

The board's most prominent feature is the Virtex-5 XC5VLX110T FPGA. The FPGA's main tool for computation is the array of 6-input look-up tables, arranged into 17280 Slices, with four look-up tables found in each Slice, adding up to a total of 69120 LUTs. A single 6-input LUT can store 64 bits of data, where its six input bits are used as an address to identify the single bit of data that is to be outputted. By manipulating the 64 bit content of the look-up table, it can be configured to carry out arbitrary Boolean functions with at most six input bits. In our design they are used to create LUTs performing a multiply-accumulate function, which are hierarchically arranged into larger and more complex modules. One out of four LUTs on the device can also be used as a 32 bit deep shift register; these are the basis to implement containers storing the data, which is directly fed to the computational module.

Attached to the board, there is a 256MB DDR2 SODIMM module, which is used for storing data exceeding the amount that can be practically stored on the FPGA.

\section{Structure of modules used in the computation}

The basic elements of the design are the LUTs denoted by $L(a,b,s)=c$, where $a,b,c$ and $s$ are two-digit binary numbers. The function carried out by $L$ is a multiply-accumulate (for short: MA) function, i.e.:
\[
c \equiv(a \cdot b)+s \mod{4} \ .
\]
Let $a=2\alpha_{1}+\alpha_{0}$, $b=2\beta_{1}+\beta_{0}$, $s=2\sigma_{1}+\sigma_{0}$, $c=2\gamma_{1}+\gamma_{0}$, where $\alpha_0,\alpha_1,\beta_0,\beta_1,\sigma_0,$ $\sigma_1,\gamma_0,\gamma_1 \in \{0,1\}$, and $L=(l_{1},l_{0})$ where $l_{1}$ and $l_{0}$ are two single bit LUTs, according to the following:
\\
\begin{itemize}
\item $l_{0}(\alpha_{0},\beta_{0},\sigma_{0})=\gamma_{0}$
\item $l_{1}(a,b,s)=\gamma_{1}$
\end{itemize}
\begin{figure}[h!]
\centering
\includegraphics[width=3.17in,height=1.2in]{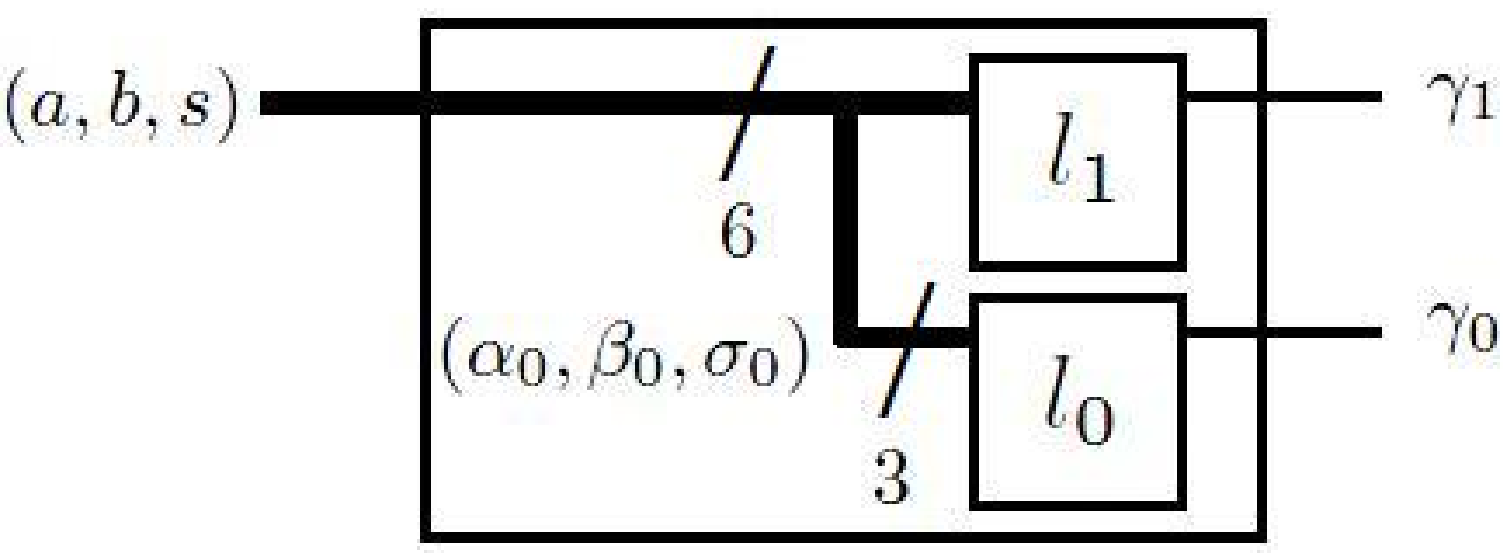}
\caption{The structure of $L(a,b,s)$}
\end{figure}

We remark that while $l_0$ needs only three input bits to accomplish its function, $l_1$ requires all six bits of input.

The LUTs $l_{0}$ and $l_{1}$ were configured to the values shown in Table 1 and Table 2 to perform the multiply-accumulate function.

\begin{table}[h!]
\centering
\begin{tabular} { | c | c | c |}
\hline
\backslashbox{($\alpha_{0},\beta_{0}$)}{$\sigma_{0}$} & 0 & 1 \\ \hline
(0,0) & 0 & 1 \\ \hline
(0,1) & 0 & 1 \\ \hline
(1,0) & 0 & 1 \\ \hline
(1,1) & 1 & 0 \\ \hline
\end{tabular}
\caption{Contents of $l_{0}$}
\end{table}

\begin{table}[h!]
\centering
\begin{tabular} { | c | c | c | c | c |}
\hline
\backslashbox{($a,b$)}{$s$} & 0 & 1 & 2 & 3 \\ \hline
(0,0) & 0 & 0 & 1 & 1 \\ \hline
(0,1) & 0 & 0 & 1 & 1 \\ \hline
(0,2) & 0 & 0 & 1 & 1 \\ \hline
(0,3) & 0 & 0 & 1 & 1 \\ \hline
(1,0) & 0 & 0 & 1 & 1 \\ \hline
(1,1) & 0 & 1 & 1 & 0 \\ \hline
(1,2) & 1 & 1 & 0 & 0 \\ \hline
(1,3) & 1 & 0 & 0 & 1 \\ \hline
(2,0) & 0 & 0 & 1 & 1 \\ \hline
(2,1) & 1 & 1 & 0 & 0 \\ \hline
(2,2) & 0 & 0 & 1 & 1 \\ \hline
(2,3) & 1 & 1 & 0 & 0 \\ \hline
(3,0) & 0 & 0 & 1 & 1 \\ \hline
(3,1) & 1 & 0 & 0 & 1 \\ \hline
(3,2) & 1 & 1 & 0 & 0 \\ \hline
(3,3) & 0 & 1 & 1 & 0 \\ \hline
\end{tabular}
\caption{Contents of $l_{1}$}
\end{table}

With the help of these basic units one can compute the dot product $w$ of two vectors $u=(u_{0},u_{1},\ldots,u_{n-1})$ and $v=(v_{0},v_{1},\ldots,v_{n-1})$. Let us define a module $m=(L[0],L[1],\ldots,L[n-1])$ by cascading $n$ MA units denoted by $L[i]$. In this module $m$ we use the output of a given MA unit as the sum input of the next unit, i.e. $s_{i+1}=c_{i}$ for $i=0,1,\ldots,n-2$, where $s_{i}$ and $c_{i}$ are the $s$ input and $c$ output of $L[i]$.
\begin{figure}[h!]
\centering
\includegraphics[width=4.5in,height=1.2in]{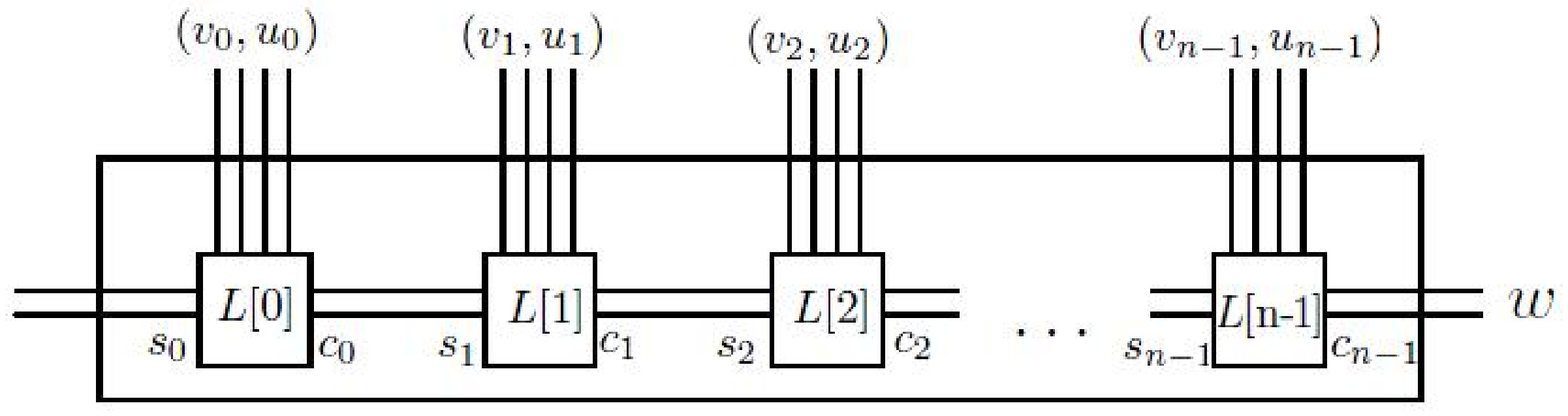}
\caption{The structure of $m(u,v)$}
\end{figure}

Therefor $m$ is a function that accepts a pair of vectors $u,v$ of two-digit numbers of length $n$ and outputs on $c_{n-1}$ the two-digit dot-product of the two vectors, i.e. $m(u,v)=w$.

In total, the number of LUTs used in $m$ is $2n$. Note that vectors of arbitrary length can be used in the computation if we connect the output of module $m$ to the sum input of $L[0]$ ($c_{n-1}=s_0$), and then iteratively shift $u$ and $v$ onto the module's input by $n$ elements at a time:
\\
\\
\\

\begin{tabbing}
Function \= $iterated\_m(u,v)$ \ \ \ // $k=length(u)=length(v)$\\
\>1. Define $\kappa=\lceil \frac{k}{n} \rceil$, $v',u' \in \mathbb{Z}_{4}^{\kappa \cdot n}$ \\
\>2. for \= $i=0$ to $\kappa \cdot n -1$ do \ \ \ // fill $v$ and $u$ with 0's\\
\> \>3. if $i<k$ then $v'_{i}=v_{i}$ else $v'_{i}=0$ \\
\> \>4. if $i<k$ then $u'_{i}=u_{i}$ else $u'_{i}=0$ \\
\>5. end for \\
\>6. Define $v_{temp},u_{temp},w$, let $w=0$ \\
\>7. for $i=0$ to $\kappa-1$ do \ \ \ // shift $v'$ and $u'$ to $v_{temp}$ and $u_{temp}$\\
\> \>8. $v_{temp}=(v'_{i \cdot n},v'_{1+(i \cdot n)},\ldots,v'_{n-1+(i \cdot n)})$ \\
\> \>9. $u_{temp}=(u'_{i \cdot n},u'_{1+(i \cdot n)},\ldots,u'_{n-1+(i \cdot n)})$ \\
\> \>10. $w=w+m(v_{temp},u_{temp})$ \\
\>11. end for \\
\>12. return $w$ \\
end Function \\
\end{tabbing}

Here $u'$ and $v'$ are the extensions of $u$ and $v$ by 0's.

We shall see that the number chosen for $n$ is critical in setting many characteristics of the entire project. The experiment used for determining $n$ will be discussed in the following chapter.

Our aim is to obtain a module that performs the matrix multiplication of $A,B\in \mathbb{Z}_{4}^{k \times k}$, where $\mathbb{Z}_{4}$ is the mod 4 residue class ring. In the following, let $C\in \mathbb{Z}_{4}^{k \times k}$ be the output matrix, such that $C=A \times B$. Furthermore, let $a_{i}$ be the $i$th row of matrix $A$ and let $b_{j}$ be the $j$th column of matrix $B$.

The multiplier units denoted by $m$ are used to create more complex modules in a hierarchical manner. First, by taking ten $m$ multiplier blocks we create a row of multipliers $R=(m_{0},m_{1},\ldots,m_{9})$. This is used to compute ten consecutive elements of a single row of the output matrix:
\[
R(a_{i},b_{j},b_{j+1},\ldots,b_{j+9})=(c_{i,j},c_{i,j+1},\ldots,c_{i,j+9})\ ,
\]
where $c_{i,j}=a_{i} \cdot b_{j}$. The input vector $a_i$ is used by all ten multiplier units of $R$. The length of these vectors, as mentioned above, can be arbitrary, but vectors of length greater than $n$ will need to be iteratively shifted to the input of $R$.

By taking ten row multipliers we can create a unit $M_{10 \times 10}=(R_{0},R_{1},\ldots,R_{9})$ which outputs a $10 \times 10$ sub-matrix of $C$:

\newpage
$M_{10 \times 10}(a_{i},a_{i+1},\ldots,a_{i+9},b_{j},b_{j+1},\ldots,b_{j+9})=$

\begin{align*}
\left(
\begin{matrix}
c_{i,j} & c_{i,j+1} & \cdots & c_{i,j+9} \\
c_{i+1,j} & c_{i+1,j+1} & \cdots & c_{i+1,j+9} \\
\vdots & \ddots \\
c_{i+9,j} & c_{i+9,j+1} & \cdots & c_{i+9,j+9} \\
\end{matrix}
\right).
\end{align*}
Finally, four such units are arranged so that a $20 \times 20$ sub-matrix of $C$ could be obtained as output:

$M_{20 \times 20}(a_{i},a_{i+1},\ldots,a_{i+19},b_{j},b_{j+1},\ldots,b_{j+19})=$
\begin{align*}
\left(
\begin{matrix}
c_{i,j} & c_{i,j+1} & \cdots & c_{i,j+19} \\
c_{i+1,j} & c_{i+1,j+1} & \cdots & c_{i+1,j+19} \\
\vdots & \ddots \\
c_{i+19,j} & c_{i+19,j+1} & \cdots & c_{i+19,j+19} \\
\end{matrix}
\right).
\end{align*}
The $M_{20 \times 20}$'s inputs are twenty vectors from both matrices $A$ and $B$.
Because of hardware constraints --- in particular the number of LUTs on the used device --- a larger arrangement of multipliers would be impractical to implement. The module $M_{20 \times 20}$ is comprised of 400 $m$ multiplier units. Figure 3 shows the hierarchy of units used to build $M_{20 \times 20}$.

\begin{figure}[h!]
\centering
\includegraphics[width=3.91in,height=1.82in]{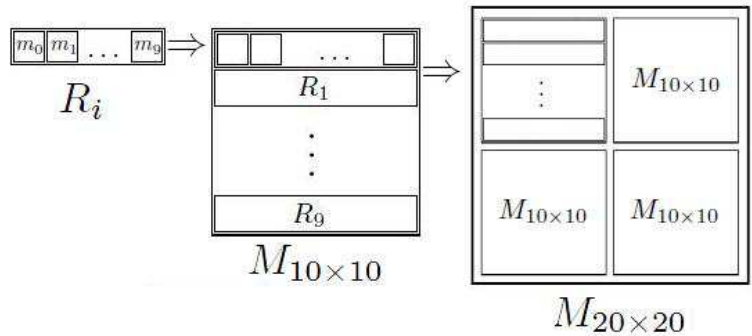}
\caption{The structure of $M_{20 \times 20}$}
\end{figure}

The $M_{20 \times 20}$ unit can be used iteratively to multiply matrices of arbitrary size, producing $20 \times 20$ sub-matrices of the output matrix $C$ with each iteration. After inputting twenty rows from matrix $A$ and twenty columns from matrix $B$ and obtaining the desired output, we can simply repeat the process for a set of rows and columns of $A$ and $B$ respectively, until we obtain the entire output matrix $C$:

\begin{tabbing}
Function \= $large\_matrix\_mult(A,B)$ \\
\>1. Define $\kappa=\lceil \frac{k}{20} \rceil$, $A',B',C' \in \mathbb{Z}_{4}^{\kappa \cdot n \times \kappa \cdot n}$ \\
\>2. for \= $i=0$ to $20\kappa -1$ do \\
\> \ \ 3. for $j=0$ to $20\kappa -1$ do \\
\> \>4. if $i<k$ and $j<k$ $a'_{ij}=a_{ij}$ else $a'_{ij}=0$ \\
\> \>5. if $i<k$ and $j<k$ $b'_{ij}=b_{ij}$ else $b'_{ij}=0$ \\
\>6. end for end for \\
\>7. for $i=0$ to $\kappa -1$ do\\
\> \ \ 8. for $j=0$ to $\kappa -1$ do \\
\> \>9. $C'[_{i+19,j+19}^{i,j}]=M_{20 \times 20}(a_{i},a_{i+1},\ldots,a_{i+19},b_{j},b_{j+1},\ldots,b_{j+19})$ \\
\>10. end for end for \\
\>11. return $C'[_{k-1,k-1}^{0,0}]$ \\
end Function \\
\end{tabbing}

Here
\[
C'[_{k,l}^{i,j}]=
\left(
\begin{matrix}
c'_{i,j} & c'_{i,j+1} & \cdots & c'_{i,l} \\
c'_{i+1,j} & c'_{i+1,j+1} & \cdots & c'_{i+1,l} \\
\vdots & \ddots \\
c'_{k,j} & c'_{k,j+1} & \cdots & c'_{k,l} \\
\end{matrix}
\right).
\]

Note that in the naive algorithm $large\_matrix\_mult(A,B)$, during the main loop (lines 7-10), for each twenty rows read from $A$, the entire matrix $B$ is read. During the whole procedure, matrix $A$ will be read entirely exactly once, while matrix $B$ will be read $\kappa$ times. Methods improving on this number are described in section 6.

Since for almost all practical cases the size $k$ of matrices $A,B\in \mathbb{Z}_{4}^{k \times k}$ will be greater than the parameter $n$, the vectors taken from these matrices will need to be iteratively shifted onto the input of the multiplier $M_{20 \times 20}$, $n$ elements at a time. Therefore, an efficient way to both store and then use the vectors taken from the matrices is the creation of FIFO type containers made of shift registers.

Let $t_{n}^{d}$ be a shift register of width $n$ and depth $d$. It means that $t_{n}^{d}$ can store at most $d$ vectors of length $n$, or equivalently a single vector of length at most $nd$. We choose $d$ such that $nd \geq k$, thus it can store one row or column from the input matrices $A$ or $B$. Let the vector filling $t_{n}^{d}$ be $f=(f_{0},f_{1},\ldots, f_{d-1})$, where $f_{i} \in \mathbb{Z}_{4}^{n}, \ i=0,1,\ldots,d-1$. In practice, $t_{n}^{d}$ is a queue data structure. In a single step, $t_{n}^{d}$ outputs a vector of length $n$ and shifts its content by $n$ places. For the $i^{th}$ activation, the container will output $f_{i}$.  After $d$ activations, the container becomes empty.

One container $t_{n}^{d}$ is used to store a single row or column of matrices $A$ or $B$ respectively. Connecting twenty of them in parallel, denoted by $T_{20n}^{d}=(t_{n}^{d}[0],t_{n}^{d}[1],\ldots,t_{n}^{d}[19])$, we obtain a container that stores twenty rows or column\-s. This is exactly the amount of data the $M_{20 \times 20}$ multiplier structure requires as input in $d$ iteration steps. After $d$ activations $T_{20n}^{d}$ has shifted all its stored data to $M_{20 \times 20}$, broken up into pieces of length $n$ for each activation. Two such $T_{20n}^{d}$ containers are connected to $M_{20 \times 20}$, one for the rows taken from matrix $A$ and one for the columns taken from matrix $B$.

\begin{figure}[h!]
\centering
\includegraphics[width=3.5in,height=1.6in]{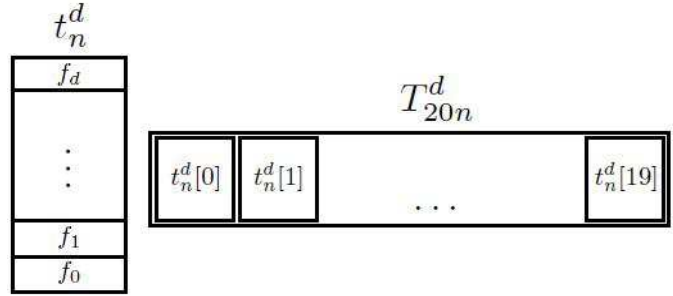}
\caption{The structure of $T_{20n}^{d}$}
\end{figure}

Using $M_{20 \times 20}$ and $T_{20n}^d$ in a proper structure, we can execute one iteration cycle of the computation. After filling one $T_{20n}^{d}$ container with the desired twenty rows from matrix $A$ and one $T_{20n}^{d}$ container with the desired twenty columns from matrix $B$, we simply send $d$ activation signals to the containers. This will shift the data onto $M_{20 \times 20}$, which computes the $20 \times 20$ product matrix in the way described in function $iterated\_m(u,v)$. The number of steps in one iteration cycle is $d$.

\section{Experimental determination of parameters}

Now, we turn to the determination of $n$ (how many MA modules should be connected into a single multiplier $m$). This sets the length of the vectors that we use in the computation in a single step and thus has an effect on many other technical parameters of the design. The goal was to find the greatest number such that the multiplier would still reliably produce the correct dot product in a single clock cycle. Clearly, this number dependents on the used hardware and the clock frequency. For the device used, the chosen clock frequency was 100 MHz, the default frequency provided by the board.

The following experiment was devised to determine the value of $n$:

Let $S$ be a multiplier $m$, called the ``Subject'', and let $E_{0},E_{1},\ldots ,E_{9}$ be ten more $m$ multipliers, called the ``Examiners''. Informally, the Examiners' duty was to verify the answers given by the Subject to questions they already knew the answer to. The ``questions'' here are test data: two vectors $v,u$ of length $n$ generated by the following sequence to obtain suitable pseudo-random values:
\[
D_{i}=D_{i-1}+D_{i-2}+2D_{i-4}+D_{i-5},
\]
where $D_{0}=D_{1}=D_{2}=D_{3}=0, \ D_{4}=1$.

More formally, let $p \in \mathbb{Z}_{10}$ be a counter that cycles between values $0,1, \ldots,9$, incrementing its value by one with each clock cycle, and returning to value $0$ after $9$. For each clock cycle during the experiment, the following happens depending on the value of $p$:
\begin{itemize}
\item The output of $S$ is checked for equality with the output of $E_{p}$. If inequality is detected, then an error is noted.
\item The test data $E_{p+1}$ is currently working on is given to $S$.
\item New test data is given to $E_{p-1}$.
\end{itemize}

\begin{tabbing}
Procedure \= testing \\
\>1. Let $S,E_{0},E_{1},\ldots,E_{9}$ be $m$ multipliers \\
\>2. Let $D$ be the test data generator \\
\>3. Let $i \in \mathbb{N}, p \in \mathbb{Z}_{10}$ \\
\>4. forever \= do \\
\> \>5. $i=i+1$ \\
\> \>6. $p \equiv i \mod{10}$ \\
\> \>7. if $S_{out} \neq E_{p\_out}$ then return ERROR \\
\> \>8. $E_{p-1\_in} \leftarrow D(i)$ \\
\> \>9. $S_{in} \leftarrow E_{p+1\_in}$ \\
\>10. end forever \\
end Procedure \\
\end{tabbing}

\begin{figure}[h!]
\centering
\includegraphics[width=4.5in,height=2.9in]{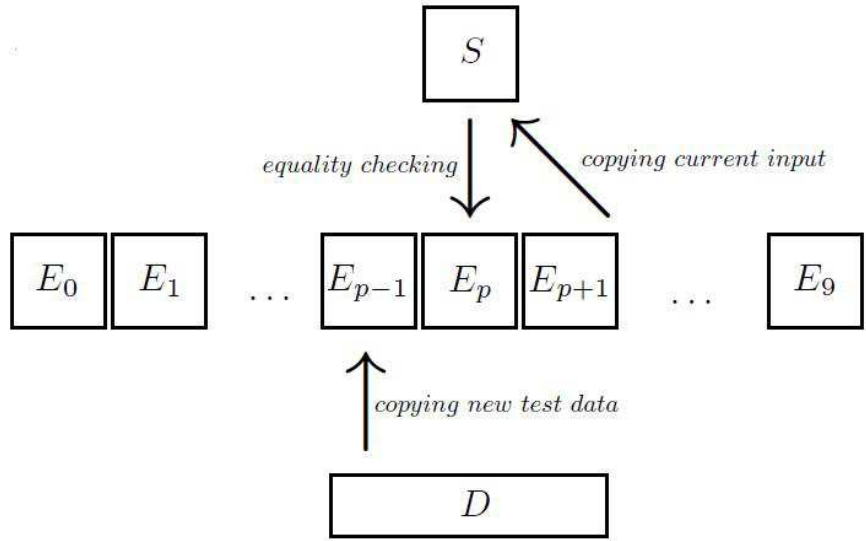}
\caption{Activity of testing module when counter's value is $p$}
\end{figure}

Note that the output of $S$ is checked every clock cycle, which yields that $S$ has only a single cycle to calculate its answer to the question it was given in the preceding clock cycle. A given Examiner, however, has ten times more time to work on its test data. Once in every ten clock cycles, new data is given to the Examiner to work on, and its output is only checked nine clock cycles later, just before it is given new input again. This way the Examiners have enough time to compute the correct answer to the question by the time it is needed.

As the initial value for $n$, we have chosen 16, a number small enough to be reasonably expected to pass the criteria set for $n$, but large enough to be of interest. If the experiment reported no error, meaning the Subject was flawlessly able to calculate the dot product for a sufficiently long time, then the value of $n$ was increased and the experiment repeated. After the first error was encountered, meaning the Subject was not able to keep up with the calculations, the largest value was chosen for $n$ for which there were no errors.

On the used device, the largest such value was found to be $28$ at a clock speed of 100 MHz and setting the length of $m$ multipliers to $n=28$ were able to work error-free for days without interruption.

\section{Computation of large matrices}

In the Section 4 we gave an algorithm for using the described modules for computing the product of large matrices. Following the description, the implemented design would make use of the parallelism offered by the FPGA only in the computation of dot products. Making further use of parallel operations, the design's performance can be significantly improved. In this section we describe the implementation choices made to raise the overall performance.

The biggest factor to consider is the management of data. When computing the product of large matrices, the amount of data to store and to move between the computation modules can easily exceed the size which can be practically stored on the FPGA. Fortunately, as mentioned before, a 256MB DDR2 SODIMM is connected to the board as the main data storage device. A module is generated using the Memory Interface Generator v3.5 intellectual property core provided by Xilinx to implement the logic needed to communicate with the DDR2 RAM. The module is structured hierarchically, connecting the memory device to a user interface. All communication with the device is done through two FIFO queues: one queue to send the command and address signals, while the other queue is used for write data and write data mask (when masking is allowed).

A naive utilization of the memory would be to simply read the required data before each iteration of the computation, and writing the output back after it is finished. An undesirable effect of this approach would be that the design would spend significantly more time with memory management than with the actual computation. The desired result would be that memory management (and all other auxiliary operations) were done during the time interval of the computation. Note that since both the size of the matrices and the multiplier module is fixed, the time the multiplication consumes is a fixed constant, which cannot be lowered. Optimally, the time of the computation should be an upper bound for the running time of the entire design. The difficulty of reaching this optimum lies in the high speed of the multiplier modules compared to the memory module.

One way to resolve the problem caused by slow transmission speed is to increase the amount of data stored on the FPGA. Informally, the main idea is to keep enough data in a prepared state, i.e. by the time the multiplier module finishes all of its computations, we have enough new data to continue working. More formally, let us define the following quantities:
\begin{itemize}
\item Let $d$ be the time necessary to complete one iteration of the computation. As described in the previous sections, this is equal to the depth of the containers $T_{20n}^{d}$.
\item Let $\kappa=\lceil \frac{k}{20} \rceil$, where $k$ is the size of the matrices. ($A,B \in \mathbb{Z}_4^{k \times k}$) This quantity is already used in algorithm $large\_matrix\_mult$. For the rest of the section, it is practical to think of $A$ and $B$ as $\kappa \times \kappa$ sized block matrices, where each element is a $20 \times 20$ matrix.
\item Let $f(A,B)$ be an arbitrary algorithm executing matrix multiplication on $A$ and $B$, including the memory management needed for the computation. Let $K(f)$ be the number of times the algorithm needs to fill a $T_{20n}^{d}$ container, i.e. the number of times it has to read twenty rows or columns from the matrices. Note that completely reading either input matrices once means filling $T_{20n}^{d}$ containers $\kappa$ times, since one $T_{20n}^{d}$ can store twenty rows or columns at a time. Algorithm $large\_matrix\_mult$'s main loop (starting at line 7) reads twenty rows from matrix $A$ (filling a $T_{20n}^{d}$ once) and reads matrix $B$ entirely for each step. Since the loop has $\kappa$ steps, it follows that $K(large\_matrix\_mult)=\kappa^2+\kappa$.
\item Let $\delta$ be the time it takes to fill a $T_{20n}^{d}$ container. This quantity depends on both the width and depth of the container. The total time $f(A,B)$ spends on reading from memory to fill the containers is $K(f)\delta$.
\item Let $\Phi(f)$ be the total time the design has to spend with memory management. This is the sum of the time it spends on reading matrices $A$ and $B$ from the memory and the time it spends on writing the product matrix $C$ into the memory. The number of times $f$ has to read $A$ and $B$ from the memory depends on $f$. Note that since the size of the total output matrix $C$ is the same as the size of $A$ and $B$, writing $C$ into the memory takes time equal to reading either matrices once from the memory. In other words, it takes $\kappa\delta$ time. The total time the design has to spend with memory management is $\Phi(f)=K(f)\delta+\kappa\delta$.
\item Let $\Gamma(f)$ be the time $f(A,B)$ spends on the computation itself. From the definition of $d$ and $\kappa$ it follows that $C(large\_matrix\_mult)=d\kappa^2$.
\end{itemize}

The goal here is to reduce $K(f)$ in such a way that the data required for the next iteration of the computation is always ready by the time the previous iteration ends. If this arrangement is achieved then $C(f)$ becomes the upper bound for the running time of the design.

Storing more data on the FPGA can be done by adding more $T_{20n}^{d}$ containers to the design. During an iteration only two such containers are used directly. The rest can be used to load data necessary for the forthcoming iteration steps.

Suppose the design has $z+2$ pieces of $T_{20n}^{d}$ containers. We assign $z$ of the containers to store rows from matrix $A$, called ``row-stores'', and two of them to store columns from matrix $B$, called ``column-stores''. With this arrangement, we can carry out $z-1$ iterations of the computation, using up the data stored in $z-1$ row-stores and one column-store. This leaves one row-store and one column-store to load new data into during the computation. Using the above definitions, the allover computation takes $(z-1)d$ time.

\begin{figure}[h!]
\centering
\includegraphics[width=4.7in,height=3.13in]{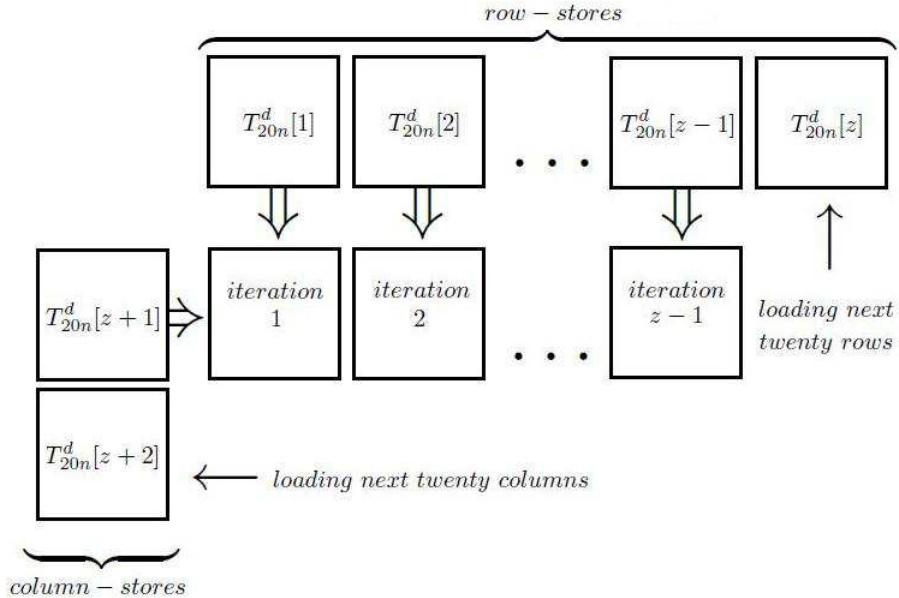}
\caption{Configuration of data stored on the FPGA}
\end{figure}

If we use all $z$ row-stores and one column-store for the computation while the remaining column-store is devoted to loading new columns into, then we would have to load all $z$ row-stores with new rows once we read all the columns before we can continue the computation. This would take $z\delta$ time for each case where we read all the columns but haven't read all the rows yet, which happens $\lfloor \frac{\kappa}{z} \rfloor$ times. In total, it would add $\lfloor \frac{\kappa}{z} \rfloor z\delta$ to the running time.

Instead, the computation of the output matrix moves slightly diagonally. See Figure 7. The $z-1$ row-stores used in the computations store a total of $(z-1) \cdot 20$ rows. Initially, the row-stores are filled with rows $a_0 \rightarrow a_{(z-1)\cdot 20 -1}$. New rows are loaded in at a slower pace than columns are. By the time all columns are read once, the contents of the row-stores have shifted exactly to the next segment of data needed, the next $(z-1)\cdot 20$ rows. After matrix $B$ is completely read once, the row-stores are filled with rows $a_{(z-1)\dot 20} \rightarrow a_{2(z-1)\cdot 20 -1}$. Reading rows and columns proceeds in this manner until we've completely read matrix $A$ once. For this reason, it is practical to choose $z$ such that $(z-1)\mid \kappa$. All together we read matrix $B$ $\frac{\kappa}{z-1}$ times and matrix $A$ once. During each $z-1$ iterations shown in Figure 6, twenty new columns and $\frac{(z-1)\cdot 20}{\kappa}$ new rows are loaded into the column-store and row-store currently unused by the computation. When the unused row-store is filled with twenty new rows, it becomes active, to be used in the following iterations. The row-store containing the rows with the least index becomes inactive in the computation and starts accepting the new rows read.

\begin{figure}[h!]
\centering
\includegraphics[width=4.7in,height=2.6in]{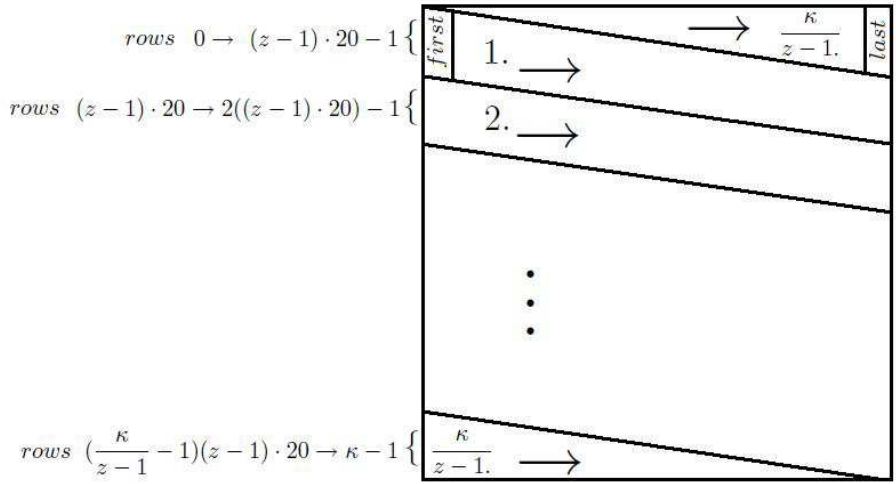}
\caption{Progression of computations through matrix $C$}
\end{figure}

\begin{tabbing}
Function \= $improved\_matrix\_mult(A,B)$ \\
\>1. Define $z,\kappa=\lceil \frac{k}{20} \rceil$, $A',B',C' \in \mathbb{Z}_{4}^{\kappa \cdot n \times \kappa \cdot n}$ \\
\>2. for \= $i=0$ to $20\kappa -1$ do \\
\>3. for $j=0$ to $20\kappa -1$ do \\
\> \>4. if $i<k$ and $j<k$ then $a'_{ij}=a_{ij}$ else $a'_{ij}=0$ \\
\> \>5. if $i<k$ and $j<k$ then $b'_{ij}=b_{ij}$ else $b'_{ij}=0$ \\
\>6. end for end for \\
\>7. Fill the row-stores with rows $a_0 \rightarrow a_{(z-1)\cdot 20-1}$ \\
\>8. Fill the column-stores with columns $b_0-b_{19}$ \\
\>9. For $i=1$ to $\frac{\kappa^2}{z-1}$ \\
\>Do in parallel:  \=$|$perform $z-1$ iterations of the computation\\
\> \> $|$READ the next 20 columns mod $\kappa \cdot 20$\\
\> \> $|$READ the next $\frac{(z-1)\cdot 20}{\kappa}$ rows mod $\kappa \cdot 20$\\
\> \> $|$WRITE the result of the previous $z-1$ iterations\\
\>11. return $C'[_{k-1,k-1}^{0,0}]$ \\
end Function \\
\end{tabbing}

The possible values for the parameters used in this section depend on the used hardware.

The size of the matrices used in the implementation are determined by parameters $n=28$ and $d=32$. The LUTs on the device that comprise the $T_{20n}^{d}$ containers can be configured as $d=32$ bit deep shift registers. For this reason the matrices are of size $896 \times 896$. Rows with length $k=896$ are the largest that can be stored in containers that are one LUT deep, making them any larger would double the number of LUTs needed for creating a $T_{20n}^{d}$. Because of the limited number of LUTs which can be used for storage purposes, $z=10$ was chosen. This yields that twelve $T_{20n}^{d}$ containers are defined in the design. Dealing with matrices larger than $k=896$ is part of future work.

For convenience, time quantities are measured in clock cycles at 100MHz, the clock speed of the $M_{20 \times 20}$ multiplier.

The value of $\delta$ depends on the DDR2 RAM used. The device was used at 200MHz, and has a 64 bit wide physical data bus.

From these values we determine the following parameters:
\begin{itemize}
\item $\kappa=\lceil \frac{896}{20} \rceil=45$,
\item $K(improved\_matrix\_mult)=\frac{\kappa}{z-1}\kappa+\kappa=\frac{45}{9}\cdot 45 + 45=270$,
\item $\delta=140$ clock cycles at 100MHz,
\item $\Phi(improved\_matrix\_mult)=K(improved\_matrix\_mult)\delta+\kappa\delta=270\cdot 140 + 45 \cdot 140=44100$ clock cycles at 100MHz,
\item $\Gamma(improved\_matrix\_mult)=d\kappa^2=32 \cdot 45^2=64800$ clock cycles at 100MHz.
\end{itemize}
The goal of $\Gamma(improved\_matrix\_mult)>\Phi(improved\_matrix\_mult)$ is a\-chieved, meaning that the running time of the design is equal to the time used by the computation.

The speedup provided by the configuration can be shown by comparing its performance to a similar implementation created on a more traditional architecture. A highly optimized C++ program was created for a machine using an Intel E8400 3GHz Dual Core processor with 2GB RAM. The algorithm is strongly specialized for the task, making use of all available options for increasing performance. It uses 64 bit long variables to perform multiplication on 16 pairs of two-digit elements at once in parallel on both processor cores.

The running time of the multiplication of matrices of the same size is over 100 ms. The FPGA implementation, as mentioned above, achieves a runtime of $\sim$0.6 ms. On average, a speedup factor of 200 is reached using the described FPGA design.

\section{Future work}

The future course of research will focus on increasing the size of the used matrices.

As mentioned in the previous section, simply increasing the depth $d$ of the $T_{20n}^{d}$ containers would be impractical. Since a single LUT on the device can only be configured as a 32 bit deep shift register, setting $d>32$ would double the number of LUTs needed for a $T_{20n}^{d}$, and the design is already using well over half of the device's LUTs that can be configured this way (13440 out of 17280, to be exact). Increasing the size of the matrices this way would require the restructuring of both the multiplier module and the algorithm used for memory management.

Instead, the currently implemented module can be used as a basic unit for the multiplication of larger matrices. Then the entries of the large matrices are $896 \times 896$ blocks.

This also allows for further optimization using Strassen's algorithm. Suppose we double the matrix sizes, interpreting them as matrices with four blocks. Using the classical algorithm, multiplying two $1792 \times 1792$ sized matrices would take eight multiplication of the blocks. Using a divide-and-conquer strategy, we can exchange one multiplication for a few extra additions.
\[
\left[
\begin{matrix}
A_{11} & A_{12} \\
A_{21} & A_{22} \\
\end{matrix}
\right]
\cdot
\left[
\begin{matrix}
B_{11} & B_{12} \\
B_{21} & B_{22} \\
\end{matrix}
\right]
=
\left[
\begin{matrix}
-D_2+D_4+D_5+D_6 & D_1+D_2 \\
D_3+D_4 & D_1-D_3+D_5-D_7 \\
\end{matrix}
\right],
\]
where
\begin{align*}
&D_1=A_{11}(B_{12}-B_{22}) \\
&D_2=(A_{11}+A_{12})B_{22} \\
&D_3=(A_{21}+A_{22})B_{11} \\
&D_4=A_{22}(B_{21}-B_{11}) \\
&D_5=(A_{11}+A_{22})(B_{11}+B_{22})\\
&D_6=(A_{12}-A_{22})(B_{21}+B_{22})\\
&D_7=(A_{11}-A_{21})(B_{11}+B_{12}).\\
\end{align*}
This algorithm, with its $O(n^{\lg  7})$ time complexity, could speed up the design on large matrices. We should note however, that the speed of the extra additions have to be carefully considered. Since the multiplication is already extremely fast, a similar improvement may also be necessary for additions if the overall performance upgrade is to remain significant.

\section*{Acknowledgements}
Research supported by the T\'AMOP 4.2.1/B-09/1/KONV-2010-0007 project and TARIPAR3 project grant Nr. TECH 08-A2/2-2008-0086.



\bigskip
\rightline{\emph{Received: May 17, 2011 {\tiny \raisebox{2pt}{$\bullet$\!}} Revised: October 11, 2011}} 

\end{document}